\documentclass[12pt]{article}
\usepackage{epsfig}

\begin{document}

\title{Nonlinear problems of complex natural systems: Sun and climate dynamics}

\author{A. Bershadskii}

\maketitle

\begin{center}
{\it ICAR, P.O. Box 31155, Jerusalem 91000, Israel}
\end{center}

\begin{abstract}
Universal role of the nonlinear one-third subharmonic resonance mechanism in generation of the 
strong fluctuations in such complex natural dynamical systems as global climate and 
global solar activity is discussed using wavelet regression detrended data. Role of the oceanic Rossby 
waves in the year-scale global temperature fluctuations and the nonlinear resonance contribution to the $El~Ni\tilde{n}o$ phenomenon have been discussed in detail. The large fluctuations of the reconstructed 
temperature on the millennial time-scales (Antarctic ice cores data for the past 400,000 years) are also shown to be dominated by the one-third subharmonic resonance, presumably 
related to Earth precession effect on the energy that the intertropical regions receive from the Sun. 
Effects of Galactic turbulence on the temperature fluctuations are discussed in this content. 
It is also shown that the one-third subharmonic resonance can be considered as a background for the 11-years solar cycle, and again the global (solar) rotation and chaotic propagating waves 
play significant role in this phenomenon. Finally, a multidecadal chaotic coherence between the detrended 
solar activity and global temperature has been briefly discussed.

\end{abstract}

\newpage

\section{Introduction}

Since the global climatology is in an earlier state of data accumulation, 
getting knowledge from the data is not an easy and straight forward process. 
However, there is only one way enabling to understand physics
of the phenomenon: accurate and punctual analysis of data. Moreover, the global climate, 
as we know it now, is a nonlinear phenomenon. This makes the process even more difficult 
and painful than usual. It is normal for this stage that the models are at best crude 
while limitations are great and (that even more dangerous) unknown.

Already the simplest energy balance model for the space averaged surface temperature $T$ has 
a form of a nonlinear equation:
$$
\frac{dT}{dt} = \frac{1}{C} \left\{ q (1-\alpha (T))- \varepsilon \sigma T^4 \right\})  \eqno{(1))}
$$
where $C$ is the heat capacity. The first term in the right hand side of the balance equation 
represents incoming solar energy with $q$ as the solar constant and $\alpha$ as the albedo, 
and second term represents outgoing infrared energy with $\varepsilon$ and $\sigma$ as the 
emissivity factor and the Stefan constant respectively. 

A more complex nonlinear model describing 
an interaction between surface energy balance and mass balance of the cryosphere has been suggested 
in Ref. \cite{sal2} (in dimensionless excess variables):
$$
\frac{ds}{dt}= \zeta  \eqno{(2)}
$$
$$
\frac{d\zeta}{dt}= c_1\zeta +c_2s -s^3-s^2\zeta+ F \sin (\omega t)  \eqno{(3)}
$$
where $s$ the sea-ice extent, $\zeta =T-s$ and T is the mean ocean surface temperature, $c_1$ and $c_2$ are certain constants, and $F \sin(\omega t)$ stands for solar forcing. It is shown in Ref. \cite{nic} that the system Eqs. (2-3) 
can be considered as a the perturbations (for $1 \gg s$) of a reference system 
$$
\frac{dx}{dt}= y  \eqno{(4)}
$$
$$
\frac{dy}{dt}=c x - x^3  \eqno{(5)}
$$ 
which is a Hamiltonian system for the conservative Duffing oscillator.

  The nonlinear Duffing oscillator appears also in the conceptual model of the 
tropical Pacific oscillations (for positive upper ocean temperature anomalies $T$ 
driving shallow thermocline depth anomalies $h$) \cite{jin},\cite{wl} 
$$
\frac{dh}{dt}= -T +\eta     \eqno{(6)}
$$

$$
\frac{dT}{dt}= -\gamma T +(h+bh^3)+ F \cos(\omega t)  \eqno{(7)}
$$
The periodic term in this equation represents the solar forcing ($\eta$ is an ambient noise, 
and $\gamma$, $b$ and $F$ are certain geophysical constants, see for more details below).

The appearance of the nonlinear Duffing oscillator in the models of the 
global climate processes for different time-space scales: from Paleoclimate Eqs. (2-5) to the decadal and inter-year oscillations Eqs. (6-7), shows that this nonlinear oscillator can be considered as a simple prototype of nonlinear climate 
systems (it will be also shown below how the nonlinear Duffing oscillator 
appears naturally in the solar dynamo models Eq. (17)). Basic symmetries of the systems can play a crucial role 
in this phenomenon (see below). For the non-linear systems far from equilibrium the oscillatory behavior can appear near bifurcation points (the dynamics generated by the Duffing oscillator can exhibit both deterministic 
and chaotic behavior \cite{ot}-\cite{b}, depending on the parameters range).

One of the specific non-linear properties of the Duffing oscillator is the so-called one-third subharmonic 
resonance \cite{nm}. Let us imagine a forced excitable system with a large amount of loosely coupled degrees of freedom 
schematically represented by Duffing oscillators with a wide range of the 
natural frequencies $\omega_0$ (it is well known \cite{tw} that oscillations with a wide range of 
frequencies are supported by ocean and atmosphere, cf. also Ref. \cite{bkb}):

$$
\ddot{x} + \omega_0^2 x +\gamma \dot{x} +\beta x^3 = F \sin\omega t    \eqno{(8)}
$$
where $\dot{x}$ denotes the temporal derivative of $x$, $\beta$ is the strength of nonlinearity, and 
$F$ and $\omega$ are characteristic of a driving force. It is known (see for instance Ref. \cite{nm}) 
that when $\omega \approx 3\omega_0$ and $\beta \ll 1$ the equation (8) has a resonant solution 
$$
x(t) \approx a \cos\left(\frac{\omega}{3}t + \varphi \right) + \frac{F}{(\omega^2-\omega_0^2)} 
\cos \omega t   \eqno{(9)}
$$
where the amplitude $a$ and the phase $\varphi$ are certain constants. 
This is so-called one-third subharmonic resonance with the driving frequency 
$\omega$. For the considered system of the oscillators an effect of synchronization can take place
and, as a consequence of this synchronization, the characteristic peaks in the spectra 
of partial oscillations coincide \cite{nl}. 
It can be useful to note, for the global climate modeling, that the odd-term subharmonic resonance 
is a consequence of the reflection symmetry of the natural nonlinear oscillators 
(invariance to the transformation $x \rightarrow -x$, cf. also Ref. \cite{mj}).

 This non-linear phenomenon can be observed directly in the real data. Let us start from the well known 
 (in relation to the global warming) graph representing 
the monthly global temperature data: figure 1 (the instrumental monthly data are available at http://lwf.ncdc.noaa.gov/oa/climate/ research/anomalies/index.html, see also Ref. \cite{s}). 
One can see that the temperature is strongly fluctuating. 
Actually, the fluctuations are of the same order as the trend itself (see figures 1 and 2). 
While the nature of the trend is widely discussed the nature of these strong fluctuations is still quite 
obscure. As we will see further these strong fluctuations have an intrinsic {\it nonlinear} nature. 
This is the one-third {\it subharmonic} resonance to annual solar forcing. Cyclic forcing, due to astronomical 
modulations of the solar input, rightfully plays a central role in the long-term climate models. 
Paradoxically, it is a very non-trivial task to find imprints of this forcing in 
the long-term climate data. It will be shown in present paper that just unusual 
properties of nonlinear response are the main source of this problem.    

  There are many well known reasons for asymmetry in response of the North and South Hemispheres 
to solar forcing: dominance of water in the Southern Hemisphere against 
dominance of land in the Northern one, topographical imbalance of land (continents) and 
oceans in the Northern Hemisphere due to continental configuration, seasonality and vegetation 
changes are much more pronounce on land than on ocean surface, and anthropogenically induced asymmetry of the last century. This asymmetry results in {\it annual} asymmetry of global heat budget and, in particular, 
in annual fluctuations of the global temperature. Nonlinear responses are expected as a 
result of this asymmetry. 
\begin{figure} \vspace{-2cm}\centering
\epsfig{width=.7\textwidth,file=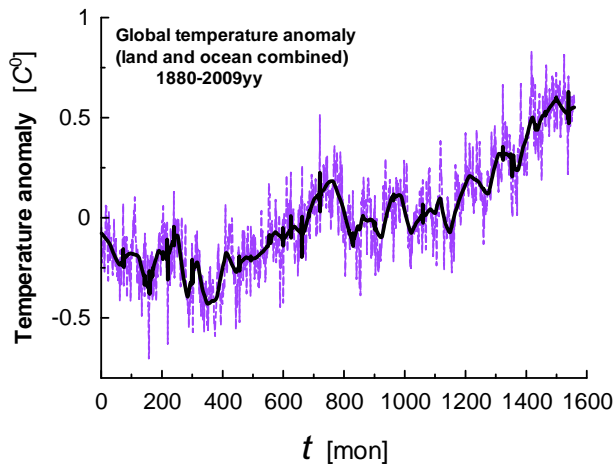} \vspace{-6cm}
\caption{The monthly global temperature data (dashed line) for the period 1880-2009. The solid curve 
(trend) corresponds to a wavelet (symmlet) regression of the data. }
\end{figure}

  The nonlinear one-third subharmonic resonance produces strong fluctuations of the global temperature 
also for much larger {\it paleoclimate} time scales: figures 3 and 4 (the reconstructed data are available at http://www.ncdc. noaa.gov/paleo/metadata/noaa-icecore-6076.html, see also Ref. \cite{ka}). Again, while the nature
of the trend is widely discussed (in relation to the glaciation cycles) the nature of these strong fluctuations 
is still quite obscure. In this case the one-third {\it subharmonic} resonance, generating the fluctuations, is presumably related to Earth precession effect on the energy that the intertropical regions receive from the Sun. 

  We will also see that in the Sun itself Nature uses the nonlinear one-third {\it subharmonic} 
resonance to amplify the hydromagnetic dynamo effects to the observed 11-year periodic solar 
activity. And again, a global 
rotation and nonlinear waves are the main components of the nonlinear mechanism. 

  Such universality of this non-linear mechanism makes it a very interesting subject for a 
thoughtful investigation. 
The problem has also a technical aspect: detrending is a difficult task for such strong fluctuations. Most of the regression methods are linear in responses. At the nonlinear nonparametric {\it wavelet} regression one chooses 
a relatively small number of wavelet coefficients to represent the underlying regression function. 
A threshold method is used to keep or kill the wavelet coefficients. At the wavelet 
regression the demands to smoothness of the function being estimated are relaxed considerably in comparison 
to the traditional methods. These advantages make the non-linear wavelet regression method an adequate tool for detrending 
the strongly fluctuating natural data.
\begin{figure} \vspace{-2cm}\centering
\epsfig{width=.7\textwidth,file=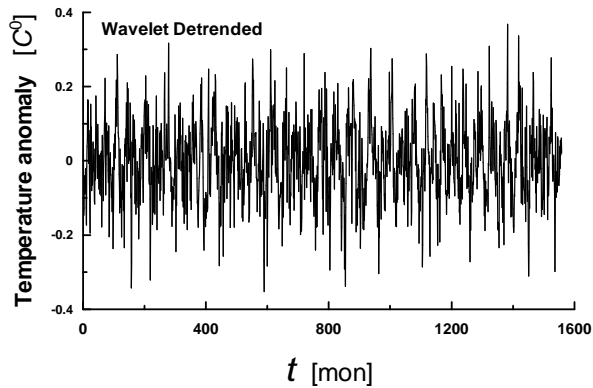} \vspace{-7cm}
\caption{The wavelet regression detrended fluctuations from the data shown in Fig. 1.}
\end{figure}

\section{One-third subharmonic resonance and Rossby waves}

Figure 1 shows (as dashed line) the instrumental monthly global temperature data 
(land and ocean combined) for the period 1880-2009, as presented at the NOAA site 
http://lwf.ncdc.noaa.gov/oa/climate/ research/anomalies/ index.html (see also Ref. \cite{s}). 
The solid curve (trend) in the figure corresponds to a wavelet (symmlet) regression of the data 
(cf. Refs. \cite{sw},\cite{o}). Figure 2 shows corresponding detrended fluctuations, which produce 
a statistically stationary set of data. We use a nonlinear nonparametric wavelet regression with 
a relatively small number of wavelet coefficients representing the underlying regression function. 
A threshold method is used to keep or 
kill the wavelet coefficients. In this case, in particular, the Universal (VisuShrink) thresholding 
rule with a soft thresholding function was used. Figure 5 shows a spectrum of the wavelet regression 
detrended data calculated using the maximum entropy method (because it provides an optimal spectral
resolution even for small data sets). One can see in this figure a small peak corresponding 
to a one-year period and a huge well defined peak corresponding to a three-years period. 
\begin{figure} \vspace{-1cm}\centering
\epsfig{width=.7\textwidth,file=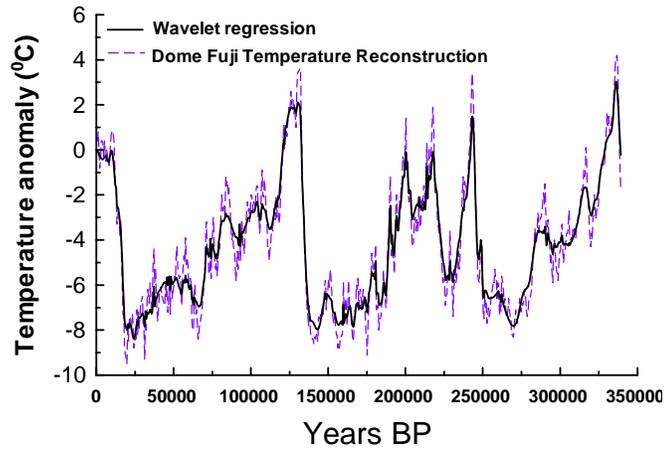} \vspace{-6cm}
\caption{The reconstructed air temperature data (dashed line) for the period 0-340 kyr. 
The solid curve 
(trend) corresponds to a wavelet (symmlet) regression of the data.}
\end{figure}
\begin{figure} \vspace{-1cm}\centering
\epsfig{width=.7\textwidth,file=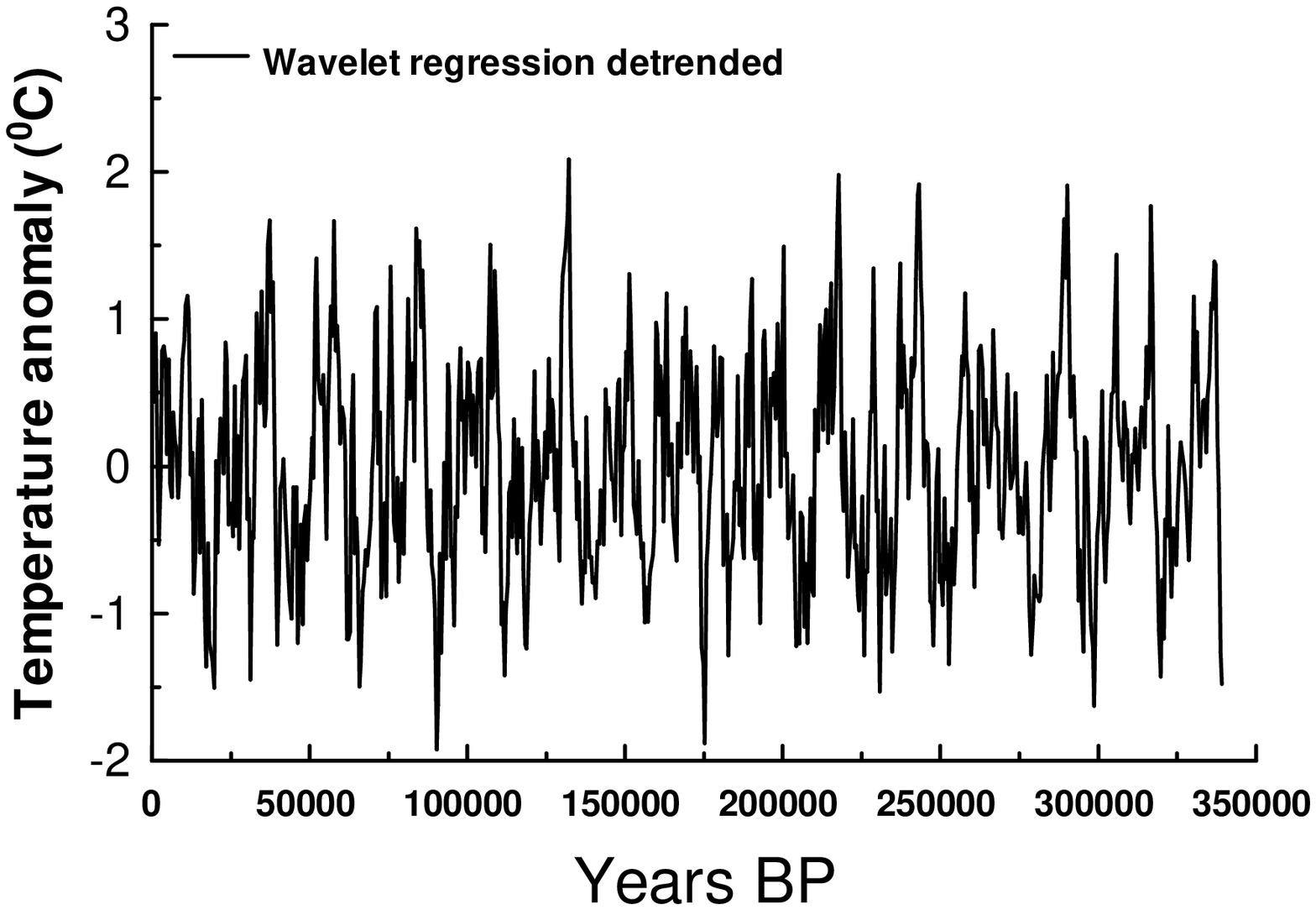} \vspace{-6cm}
\caption{The wavelet regression detrended fluctuations from the data shown in Fig. 3. }
\end{figure}

\begin{figure} \vspace{-2cm}\centering
\epsfig{width=.7\textwidth,file=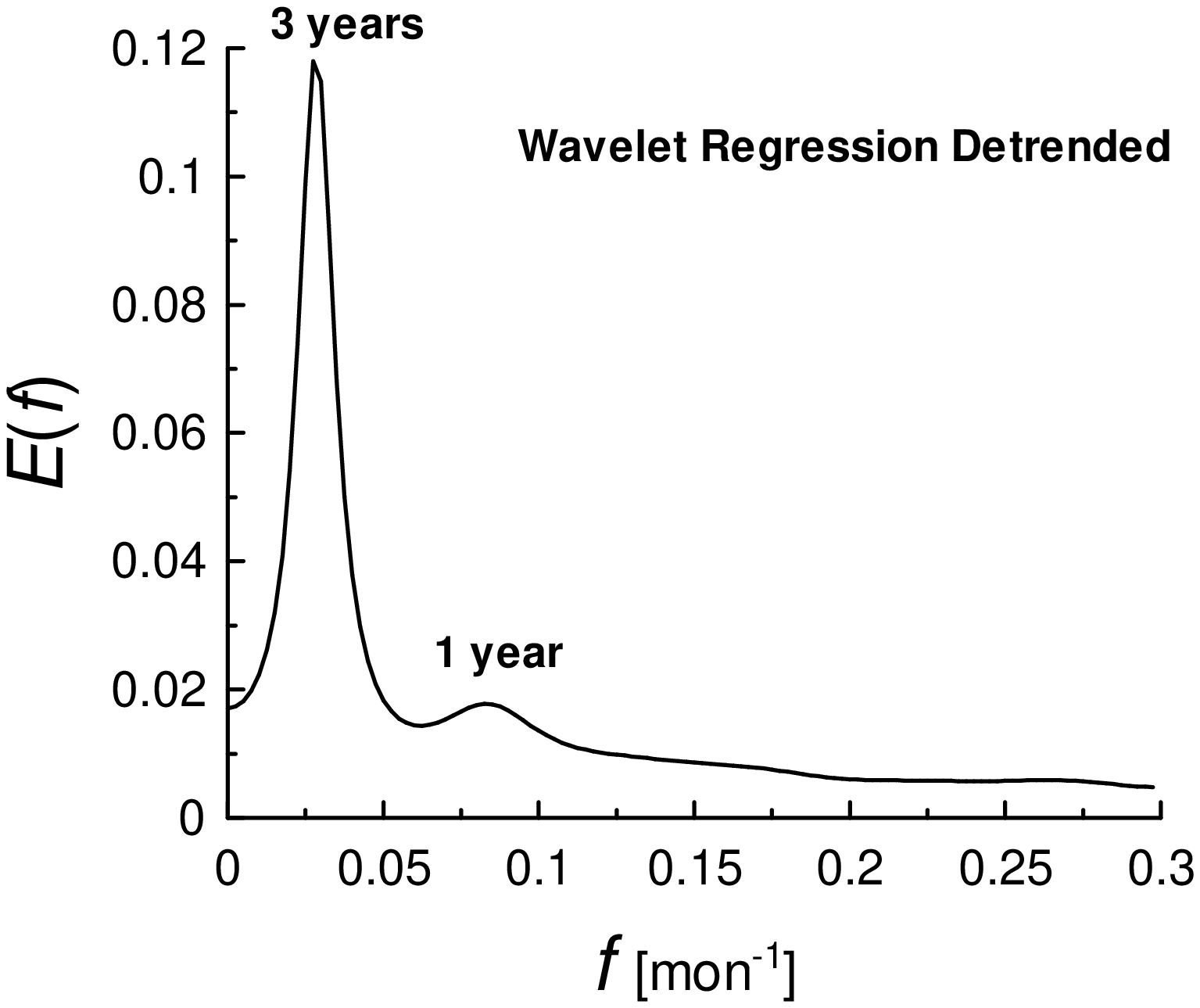} \vspace{-5.5cm}
\caption{Spectrum of the wavelet regression detrended fluctuations of 
the monthly global temperature anomaly (land and ocean combined).}
\end{figure}

This is the one-third subharmonic resonance with the driving frequency 
$\omega$ corresponding to the {\it annual} NS-asymmetry of the solar forcing 
(the huge peak in Fig. 5 corresponds to the first term in the right-hand side of the Eq. (9)). 

\begin{figure} \vspace{-2cm}\centering
\epsfig{width=.7\textwidth,file=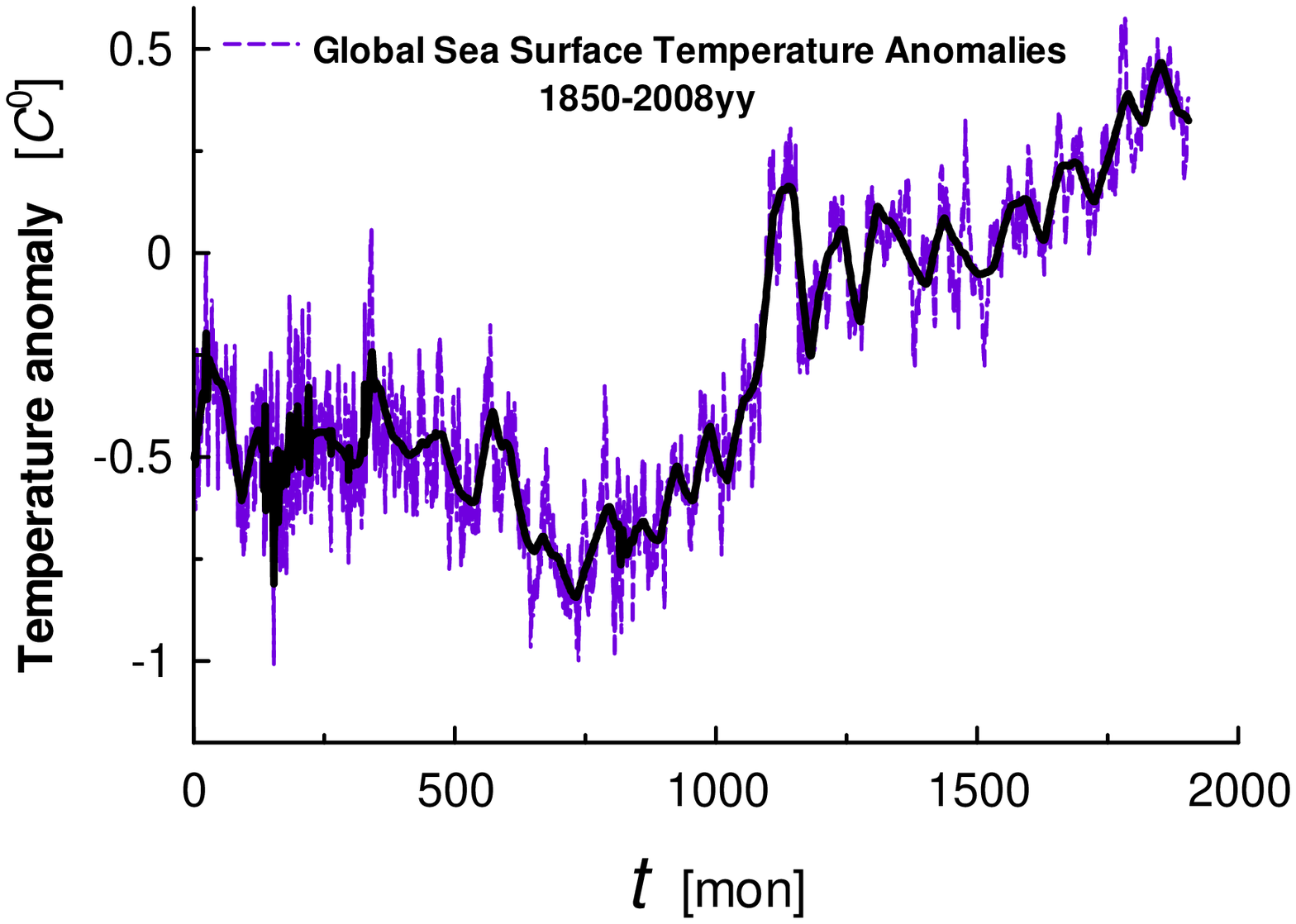} \vspace{-6cm}
\caption{The monthly global Sea Surface Temperature Anomalies (dashed line) for the period 1850-2008yy. 
The solid curve (trend) corresponds to a wavelet 
(symmlet) regression of the data.}
\end{figure}
\begin{figure} \vspace{-2cm}\centering
\epsfig{width=.7\textwidth,file=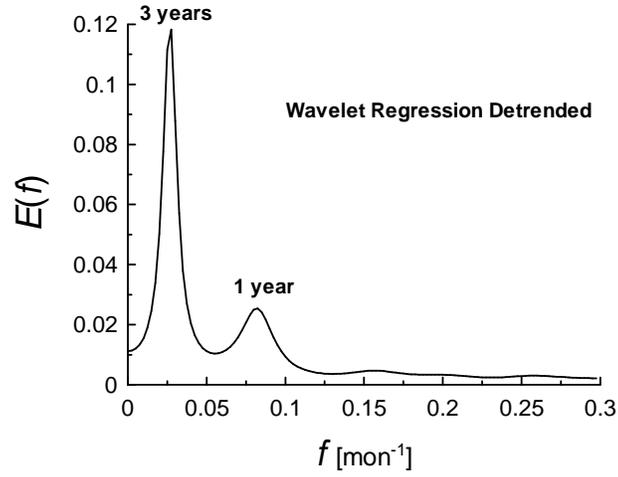} \vspace{-5.5cm}
\caption{The same as in Fig. 5 but for the wavelet regression detrended fluctuations of the global Sea Surface Temperature Anomalies.}
\end{figure}

\begin{figure} \vspace{-0.5cm}\centering
\epsfig{width=.7\textwidth,file=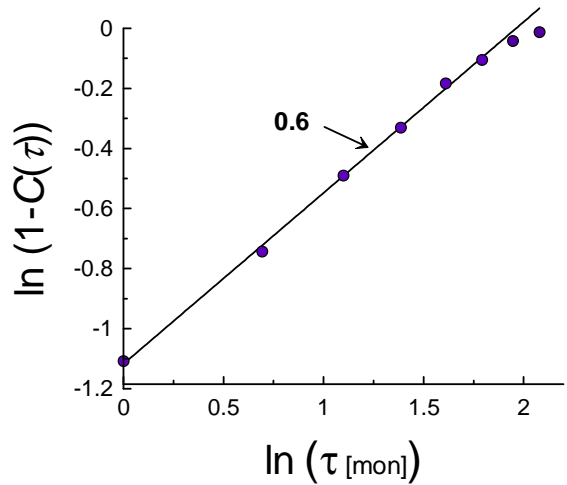} \vspace{-5cm}
\caption{Defect of autocorrelation function versus $\tau$ (in ln-ln scales) for the wavelet regression detrended fluctuations of the global Sea Surface Temperature Anomalies. The straight line is drawn in order to indicate 
scaling Eq. (10). }
\end{figure}
\begin{figure} \vspace{-1cm}\centering
\epsfig{width=.5\textwidth,file=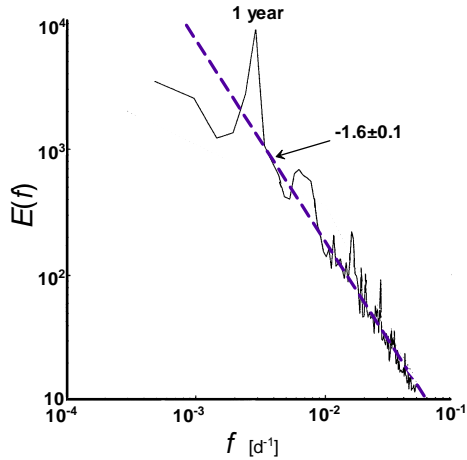} \vspace{-2cm}
\caption{Spectrum of sea surface height fluctuations \cite{zw} (TOPEX/
Poseidon and ERS-1/2 altimeter measurements) in the logarithmic scales. 
The profound peak corresponds to the annual cycle. 
The dashed straight line is drawn in order to indicate correspondence to the scaling Eq. (10).}
\end{figure}
\begin{figure} \vspace{-0.5cm}\centering
\epsfig{width=.7\textwidth,file=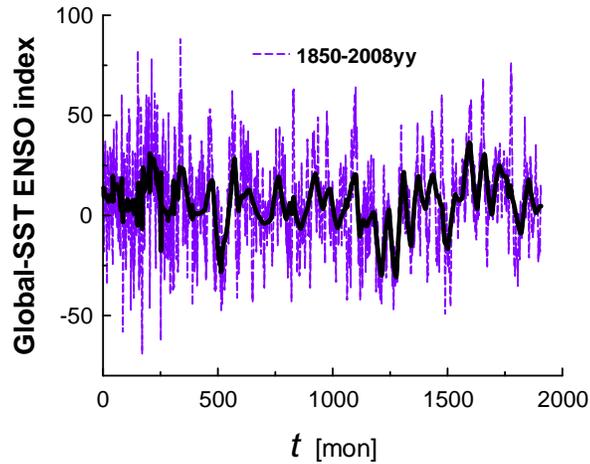} \vspace{-6cm}
\caption{The monthly Global-SST ENSO index (dashed line) for the period 1850-2008yy (the index is in hundredths 
of a degree Celsius). The solid curve (trend) corresponds to a wavelet (symmlet) regression of the data. }
\end{figure}
\begin{figure} \vspace{-1.5cm}\centering
\epsfig{width=.7\textwidth,file=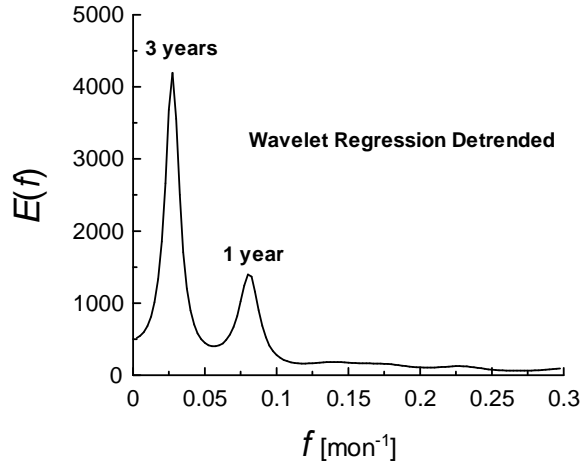} \vspace{-5.5cm}
\caption{The same as in Fig. 5 but for the wavelet regression detrended fluctuations 
of the Global-SST ENSO index. }
\end{figure}
The fluctuations of oceanic temperature cause certain variations of the sea surface height. 
These variations are intermixed with the sea surface height variations caused by the oceanic planetary Rossby waves. 
The oceanic planetary Rossby waves play an important role in the response of the global 
ocean to the forcing (see, for instance, Refs. \cite{pl},\cite{zw}) and they are of fundamental importance 
to ocean circulation on a wide range of time scales (it was also suggested that the Rossby 
waves play a crucial role in the initiation and termination of the $El~Ni\tilde{n}o$ phenomenon, see also below). 
Therefore, they present a favorable physical background for the global subharmonic resonance. It should be noted 
that the variable $h$ in equation (7) represents western boundary Rossby wave
reflection of its counterpart in the equation (6). That provides a
negative feedback to the $T$ tendency in the eastern equatorial ocean via a non-linear thermocline displacement: 
$h+bh^3$ (its vertical gradient separating near-surface and deep-ocean mixed
layers \cite{jin},\cite{wl}). The solar periodic forcing plays analogous role in this equation.

For that reason it is interesting to look also separately at global sea surface temperature anomalies. 
These data for time range 1850-2008yy are shown in Fig. 6 (the monthly data are available at 
http://jisao.washington.edu/data/global sstanomts/, see also Ref. \cite{s2}). The solid curve (trend) in the figure 
corresponds to the wavelet (symmlet) regression of the data.  
Figure 7 shows a spectrum of the wavelet regression detrended data 
calculated using the maximum entropy method. The spectrum seems to be 
very similar to the spectrum presented in Fig. 5. 
\begin{figure} \vspace{-1.5cm}\centering
\epsfig{width=.7\textwidth,file=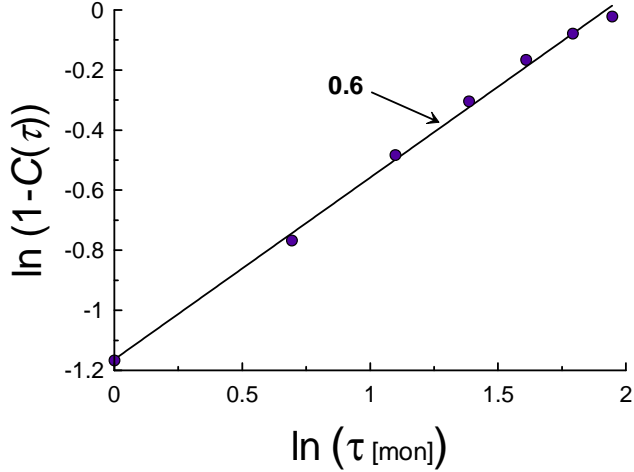} \vspace{-5cm}
\caption{The same as in Fig. 8 but for the wavelet regression detrended fluctuations 
of the Global-SST ENSO index.}
\end{figure}
Since the high frequency part of the spectrum is corrupted by strong fluctuations (the Nyquist frequency 
equals 0.5 [$mon^{-1}$]), it is interesting to look at corresponding autocorrelation function $C(\tau)$ in order to understand what happens on the monthly scales. It should be noted that scaling of defect of the autocorrelation 
function can be related to scaling of corresponding spectrum:
$$
1-C(\tau) \sim \tau^{\alpha}~~~~~~\Leftrightarrow~~~~~~~~ E(f) \sim f^{-(1+\alpha)}   \eqno{(10)}
$$
Figure 8 shows the defect of the autocorrelation function in ln-ln scales in order to estimate the scaling 
exponent $\alpha \simeq 0.6 \pm 0.04$ (the straight line in this figure indicates the scaling Eq. (10)). 
The existence of oceanic Rossby waves was confirmed rather recently by NASA/CNES TOPEX/Poseidon satellite 
altimetry measurements. Corresponding to these measurements spectrum of the sea surface height 
fluctuations, calculated in Ref. \cite{zw} with {\it daily} resolution (see also \cite{zfw}), 
is shown in Figure 9. The dashed straight line in this figure is drawn in order to indicate 
correspondence to the scaling Eq. (10): $1+\alpha \simeq 1.6$.

The Rossby waves (together with Kelvin waves) and a strong atmosphere-ocean feedback provide physical 
background for the $El~Ni\tilde{n}o$ phenomenon (see, for instance, Refs. \cite{jin},\cite{wl},\cite{tzi} and references therein). 
Figure 11 shows spectrum for the wavelet detrended fluctuations 
of the so-called Global-SST ENSO index (Fig. 10), which captures the low-frequency part 
of the $El~Ni\tilde{n}o$ phenomenon (the monthly data are available at http://jisao.washington.edu/
data/globalsstenso/). The annual forcing can come from the oceanic Rossby waves (cf Fig. 9). 
To support this relationship we show in figure 12 defect of autocorrelation function calculated using the 
wavelet detrended fluctuations from Fig. 10. 
The ln-ln scales have been used in Fig. 12 in order to estimate the scaling 
exponent $\alpha \simeq 0.6 \pm 0.03$ (the straight line in this figure indicates the scaling Eq. (10), 
cf. Figs. 8 and 9). Using these observations one can suggest that the $El~Ni\tilde{n}o$ phenomenon 
has the one-third subharmonic resonance as a background. 
\begin{figure} \vspace{-2cm}\centering
\epsfig{width=.7\textwidth,file=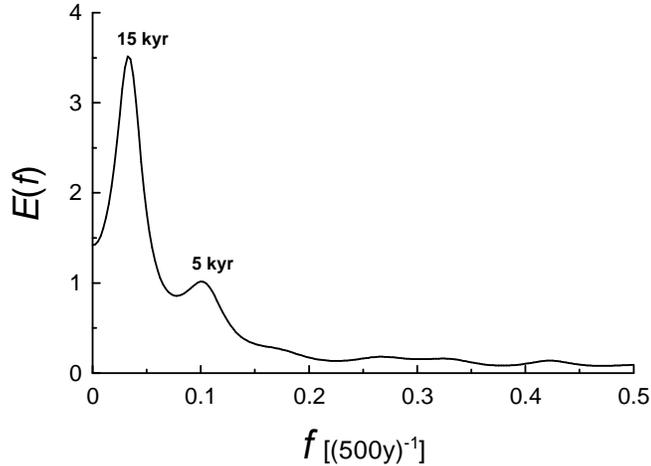} \vspace{-5.5cm}
\caption{Spectrum of the wavelet regression detrended fluctuations shown 
in Fig. 4.}
\end{figure}
\section{Nonlinear Paleoclimate}

Recent paleoclimate reconstructions provide indications of nonlinear properties of 
Earth climate at the late Pleistocene \cite{rh},\cite{sal} (the period from 0.8 Myr to present). 
Long term decrease in atmospheric $CO_2$, which could result in a change in the internal 
response of the global carbon cycle to the obliquity forcing, 
has been mentioned as one of the principal reasons for this phenomenon (see, for instance, 
\cite{berg1}-\cite{clar}). At present time one can recognize at least two problems 
of the nonlinear paleoclimate, which we will address in present paper using recent data and 
speculations \cite{b3}. 

{\bf A}.  Reconstructed air temperature  on millennial time scales are known 
to be strongly fluctuating. See, for instance figures 3 and 4.  While the nature of the trend 
is widely discussed (in relation to the glaciation cycles) the nature of these strong fluctuations 
is still quite obscure. The spectral analysis of the wavelet regression detrended data reveals 
rather surprising nature of the strong temperature fluctuations. Namely, the detrended fluctuations 
of the reconstructed temperature are completely dominated by the one-third subharmonic resonance, 
presumably related to  Earth precession effect on equatorial insolation. 

{\bf B}. Influence of Galactic turbulent processes on the Earth climate can be very significant for time-scales 
less than 2.5 kyr. \\

Figure 3 shows reconstructed air temperature data (dashed line) for the period 0-340 kyr 
as presented at the site http://www.ncdc.noaa.gov/paleo/ metadata/noaaicecore-6076.html 
(Antarctic ice cores data, see also Ref. \cite{ka}). 
The solid curve (trend) in the figure corresponds to a wavelet (symmlet) regression of the data 
(cf. Ref. \cite{o}). Figure 4 shows corresponding detrended fluctuations, which produce 
a statistically stationary set of data. Figure 13 shows a spectrum of the wavelet regression detrended data 
calculated using the maximum entropy method. One can see in this figure a small peak corresponding 
to period $\sim$ 5kyr and a huge well defined peak corresponding to period $\sim$15kyr. 
We also obtained analogous results (approximately 10\% larger) from the "Vostok" ice core data 
for period 0-420kyr (for the data description see Refs. \cite{vostok1},\cite{vostok2}).

Origin of the periodic energy input with the period $\sim$ 5kyr can be related 
to dynamics of the energy that the intertropical regions receive 
from the Sun (equatorial insolation). Indeed, it is found in Ref. \cite{blm} that 
a clear and significant 5kyr period is present in this dynamic over last 1 Myr. 
The amplitude of the 5kyr cycle in the insolation decreases rapidly when getting 
away from the equator. Using the fact that double insolation maximum and minimum arise 
in the tropical regions in the course of one year, the authors of the Ref. \cite{blm} 
speculated that this period in seasonal amplitude of equatorial insolation is determined 
by fourth harmonic of the Earth precession cycle. It should be noted, that the idea of a significant 
role of tropics in generating long-term climatic variations is rather a new one (see Ref. \cite{blm} 
for relevant references). In Ref. \cite{hag}, for instance, the authors speculated that the 
high frequency climate variability (in the millennial time scales) could be related to high 
sensitivity of the tropics to summer time insolation. Then, the oceanic advective transport 
could transmit an amplified response of tropical precipitation and temperature to high latitudes. 
Physical mechanism of this amplification is still not clear and the above discussed one-third subharmonic 
resonance can be a plausible possibility (in this respect it is significant, that we used the 
Antarctic data).  

\section{Galactic turbulence and the temperature fluctuations}
\begin{figure} \vspace{-2cm}\centering
\epsfig{width=.7\textwidth,file=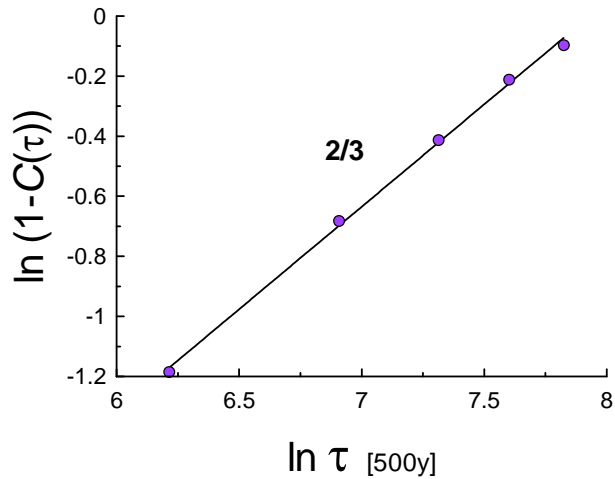} \vspace{-5.5cm}
\caption{A small-time-scales part of the autocorrelation function defect of the wavelet regression
detrended fluctuations from the data shown in Fig. 4. The straight
line indicates the Kolmogorov's '2/3' power law for the structure function.}
\end{figure}

Since the high frequency part of the spectrum is corrupted by strong fluctuations (the Nyquist frequency 
equals 0.5 [$(500y)^{-1}$]), it is interesting to look at corresponding autocorrelation function $C(\tau)$ 
and at structure functions $S_p(\tau)$ (of different orders $p$) in order to understand what happens on 
the millennial time scales. The correlation function defect $1- C(\tau )$ is proportional to the second order structure function $S_2(\tau)$. Therefore, we can compare results obtained by these different tools. First let us look at autocorrelation function $C(\tau)$. Figure 14 shows a relatively small-times part of the correlation function defect. The ln-ln scales have been used in this figure in
order to show a power law (the straight line) for the second order structure function: $S_2 (\tau) = \langle |x(t+\tau)- x(t)|^2 \rangle$ :
$$
1- C(\tau ) \propto  S_2(\tau) \propto \tau^{2/3}  \eqno{(11)}
$$
This power law: '2/3', for structure function (by virtue of the Taylor hypothesis
transforming the time scaling into the space one \cite{my},\cite{b1})
is known for fully developed turbulence as Kolmogorov's power law.

Although, the scaling interval is short, the value of the exponent is rather intriguing.
This exponent is well known in the theory of fluid (plasma) turbulence and corresponds to so-called
Kolmogorov's cascade process. This process is very universal for turbulent fluids
and plasmas \cite{gibson},\cite{pet}. In spite of the fact that magnetic field is presumably important for
interstellar turbulence the Kolmogorov description can be still theoretically acceptable even in this area
\cite{l}-\cite{clv}. Moreover, the Kolmogorov-type spectra were observed on the
scales up to kpc. In order to support the Kolmogorov turbulence as a background of the
wavelet regression detrended temperature modulation we calculated also structure functions
$S_p (\tau) = \langle |x(t+\tau)- x(t)|^p \rangle$ with different orders $p$. In the classic Kolmogorov
turbulence (at very large values of the Reynolds number \cite{my},\cite{sd})
$$
S_p \propto \tau^{\zeta_p}~~~~~~~ \zeta_p \simeq \frac{p}{3}   \eqno{(12)}
$$
(at least for $p \leq 3$). Figure 15 shows a small-time-scales part of the structure functions $S_p$ with $p=0.2,~0.5,~0.7,~1,~2,~3$
for the wavelet regression detrended fluctuations from the data shown in Fig. 4. The straight lines
are drawn in order to indicate scaling in the ln-ln scales. Figure 16 shows as circles the scaling exponent
$\zeta_p$ against $p$ for the scaling shown in Fig. 15 (the exponents were calculated using slopes of the
straight lines - best fit, in Fig. 15). The bars show the statistical errors. The straight line in Fig. 16
corresponds to the strictly Kolmogorov turbulence with $\zeta_p =p/3~$ Eq. (12). One can see good agreement with the
Kolmogorov turbulence modulation.
\begin{figure} \vspace{-2cm}\centering
\epsfig{width=.7\textwidth,file=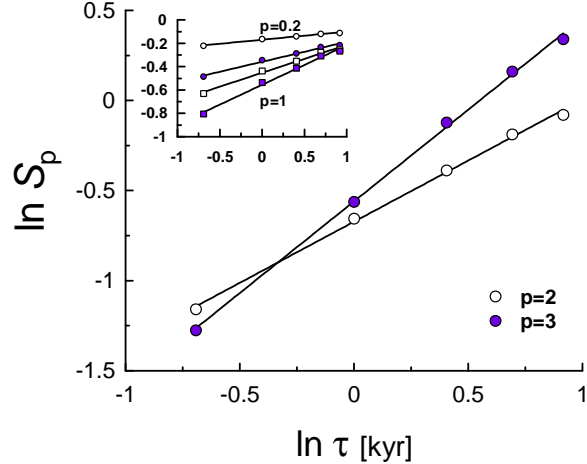} \vspace{-5.5cm}
\caption{A small-time-scales part of the structure functions $S_p$ (p=2,3) for the wavelet regression
detrended fluctuations from the data shown in Fig. 4. The insert shows the structure functions for
$p=0.2,~0.5,~0.7,~1$. The straight line indicates scaling in the ln-ln scales.}
\end{figure}

For turbulent processes on Earth and in Heliosphere
the Kolmogorov's scaling with such large time-scales certainly cannot exist. Therefore, one should think about
a Galactic origin of Kolmogorov turbulence (or turbulence-like processes \cite{g2}) with such large time-scales. 
Let us recall that diameter of the Galaxy is approximately 100,000 light years.
This is not surprising if we recall possible role of the
galactic cosmic rays for Earth climate.  Galactic cosmic ray
intensity at the Earth's orbit is modulated by galactic turbulence \cite{b1}. On the other hand,
the galactic cosmic rays can determine the amount of cloud cover (a very significant climate factor)
on global scales through the massive aerosols formation (see, for instance, \cite{sf}-\cite{du}). Thus,
the galactic turbulence can modulate the global temperature fluctuations by the Kolmogorov
scaling properties on the millennial time scales. 
\begin{figure} \vspace{-0.5cm}\centering
\epsfig{width=.7\textwidth,file=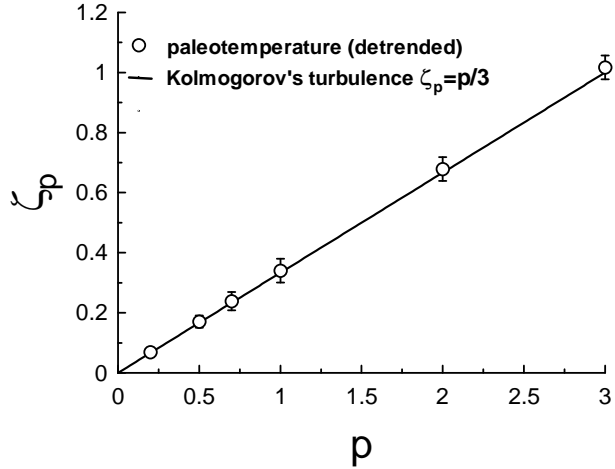} \vspace{-6cm}
\caption{The scaling exponent $\zeta_p$ against $p$ for the scaling shown in Fig. 15 (circles).
The straight line is drawn in order to indicate the Kolmogorov's turbulence with $\zeta_p =p/3$ Eq. (12). }
\end{figure}
If one knows the characteristic velocity scale $v$ for the Taylor hypothesis one can estimate 
outer space-scale of corresponding galactic turbulence as $L \geq 2500y \times v$. 
However, it is not clear what estimate we should take for the $v$. For instance, one could try velocity of the 
solar system relative to the cosmic microwave background (CMB) rest frame: $v \sim 370$ km/sec. In this case 
one obtains $L \sim 1$pc. It should be noted that in recent paper \cite{h} it is suggested that the 
typical outer scale for spiral arms can be as small as 1pc and in interarm regions the outer scale can be larger 
than 100pc. Since the solar system and Earth are at present time within the Orion Arm this suggestion is in 
agreement with the above estimate. Although the suggestion of the Ref. \cite{h} is still under active discussion 
the paleoclimate consequences of this suggestion can be very interesting and we will discuss one of them here. 
Namely, when orbiting the Galactic center the solar system and Earth are in the interarm regions the 
reverse of the Taylor hypothesis provides us with the outer time-scale $\sim 2500 \times 100$ years. This time 
scale is larger then any known glaciation period (which are determined by the periods related to orbiting 
Earth around Sun, see for instance \cite{b}). Strong {\it turbulent} fluctuations of the cosmic rays flux on such 
large time-scales should prevent to the glaciation cycles to occur when the solar system is in the 
interarm regions. The Earth deglacitaion related to the interarm regions was suggested in Refs. 
\cite{shav1},\cite{shav2} and explained by difference in intensity of the cosmic ray flux in the spiral arms 
and in the interarm regions. It is difficult to estimate at present time which of the two mechanisms is 
more efficient. In any way they are working to the same end and both are open to discussion. 

\section{Subharmonic resonance in solar activity}

The solar activity is chaotic but has a well-defined mean period of
about 11 years. The 11-year cycle is well known for more than a century and a half. 
Despite this, nature of the 11-year cycle is still a subject of vigorous investigations. 
The most popular point of view is a 'dynamo-wave' mechanism. It is assumed that a magnetic 
dynamo, generated by the solar differential rotation and the helicity of turbulent convective
flows, produces this propagating wave. Helicity, through the $\alpha$-effect, plays 
crucial role in this mechanism. Recently, observational 
data of physical quantities associated with the $\alpha$-effect became available and a 
considerable progress in this direction was achieved (see, for instance, Refs. 
\cite{kle},\cite{zha},\cite{sokol}). 
The $\alpha$-effect has two contributors: one related to helicity of convective 
vortices and another related to the helicity of magnetic field. Intrinsically non-linear character 
of the problem makes it especially difficult for theoretical investigation. The nonlinear 
solar dynamo has to be saturated in order to get a quasi-stationary wave. The magnetic part 
of the $\alpha$-effect can play a crucial role in such saturation, while certain modification 
of the turbulent diffusivity and other transport coefficients is unavoidable at this 
process (see Refs. \cite{kle},\cite{zha},\cite{sokol} and section 'Chaotic dynamo'). 
Results of the above mentioned simulations show that the dynamo model leads to a
steadily oscillating magnetic configuration. The cyclic behavior is typical for
moderate dynamo action whereas for the stronger dynamo action a chaotic behavior is usually observed 
in the dynamo simulations. 
\begin{figure} \vspace{-1cm}\centering
\epsfig{width=.7\textwidth,file=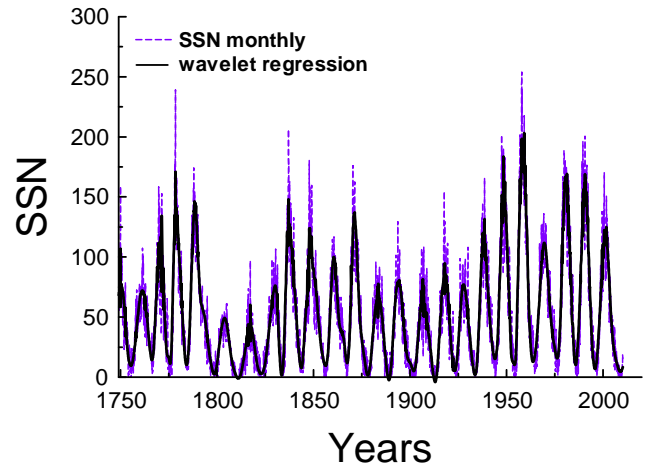} \vspace{-5.5cm}
\caption{The monthly sunspot number (dashed line) for the period 1749-2009 years. 
The solid curve 
(trend) corresponds to a wavelet (symmlet) regression of the data. }
\end{figure}
\begin{figure} \vspace{-1cm}\centering
\epsfig{width=.7\textwidth,file=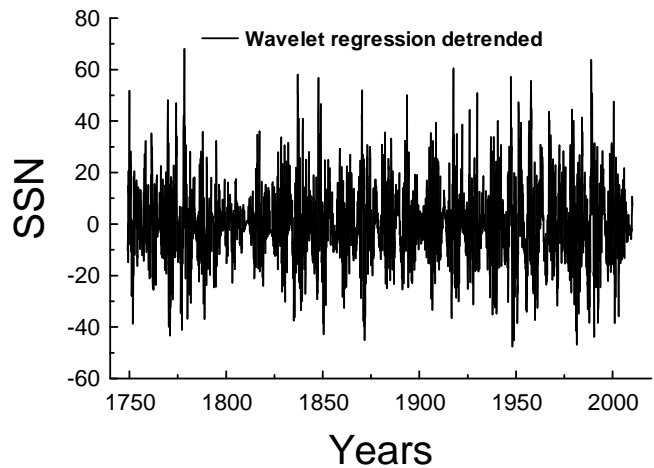} \vspace{-6cm}
\caption{The wavelet regression detrended fluctuations from the data shown in Fig. 17.}
\end{figure}
  It is believed that the dynamo mechanism is mainly operational deep inside 
the convective zone (or even in the overshoot layer). However, in the upper layers of the convection 
zone ('at surface') a strong hydromagnetic activity has been also observed. The presence
of large-scale meandering flow fields (like jet streams), banded zonal flows and evolving meridional 
circulations produces a very complex picture \cite{b4}. In this situation one can expect that more than one non-linear 
mechanisms can be at background of the observed 11-year solar cycle. Namely, it could be inferred from 
the data analysis presented below that a dynamo mechanism, directly affected by the solar rotation, 
may generate the basic chaotic oscillations (with a fundamental period different from the 11-year cycle), 
while another non-linear mechanism amplifies these oscillations to the observed 11-year chaotic oscillations. 
In this two-stage picture the mirror asymmetry of the solar magnetic field can be still a key driver of the 11-year activity cycle through an additional non-linear amplifying mechanism.\\

Most of the regression  methods are linear in responses and statistical analyses of the experimental sunspot 
data was dominated by linear stochastic methods, while it was recently rigorously 
shown in Ref. \cite{pn} that a nonlinear dynamical mechanism (presumably a driven nonlinear 
oscillator, see also Ref. \cite{mini}) determines the sunspot cycle. Figure 17 shows 
the monthly sunspot number (dashed line) for the period 1749-2009 years 
(the data are available at http://sidc.oma.be/sunspot-data/). The solid curve 
(trend) corresponds to a wavelet (symmlet) regression of the data (cf. Refs. \cite{o}). 
Figure 18 shows corresponding detrended fluctuations, which produce 
a statistically stationary set of data. The wavelet detrended data set is not statistically 
stationary, however, for the extended 1611-2009 period, presumably due to the Maunder minimum 
in the 17th century (see Refs. \cite{frick},\cite{sokol2}). 
Figure 19 shows a spectrum of the wavelet regression detrended data 
calculated using the maximum entropy method (because it provides an optimal spectral
resolution even for small data sets). In Fig. 19 one can see a well defined peak corresponding 
to period $\sim$ 3.7 years.

The wavelet regression method 
detrends the data from the approximately 11-years period (cf. Fig. 17). Therefore, it is plausible 
that the one-third subharmonic resonance \cite{nm} can be considered as a background for 
the 11-years solar cycle: $11/3.7 \simeq 3$. Indeed, it is known \cite{nocera} that 
interaction of the Alfven waves (generated in a highly magnetized plasma by a cavity's moving boundaries) 
with slow magnetosonic waves can be described using Duffing oscillators (see also Refs. \cite{pn},\cite{plo} and below).

\section{Chaotic dynamo}
\begin{figure} \vspace{-1cm}\centering
\epsfig{width=.7\textwidth,file=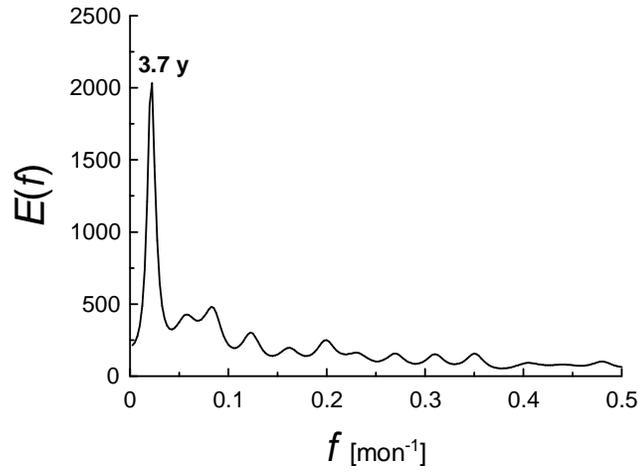} \vspace{-6cm}
\caption{Spectrum of the wavelet regression detrended fluctuations shown 
in Fig. 18.}
\end{figure}
\begin{figure} \vspace{-1cm}\centering
\epsfig{width=.7\textwidth,file=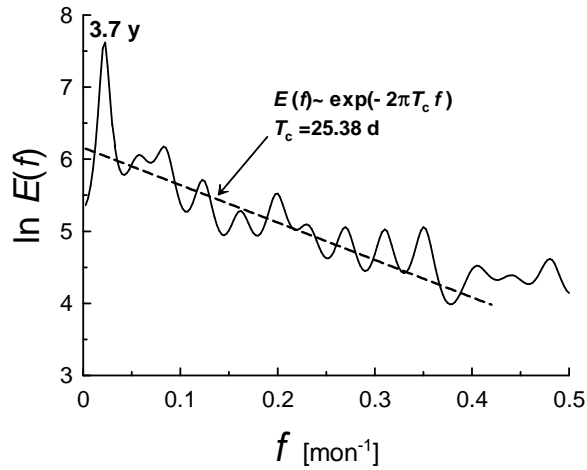} \vspace{-5.5cm}
\caption{The same as in Fig. 19 but in semi-logarithmical scales. The dashed straight line 
indicates an exponential decay.}
\end{figure}
In order to understand appearance of the $3.7$-years period let us represent the spectrum shown in Fig. 19 
in semi-logarithmical scales: figure 20. In these scales an exponential behavior corresponds to a straight 
line. It is known, that both stochastic and deterministic processes can result in the
broad-band part of the spectrum, but the decay in the
spectral power is different for the two cases. The exponential
decay indicates that the broad-band spectrum
for these data arises from a deterministic rather than a
stochastic process. For a wide class of deterministic
systems a broad-band spectrum with exponential
decay is a generic feature of their chaotic solutions 
Refs. \cite{oht}-\cite{fm}. A wavy exponential decay (see Fig. 20) is a characteristic of 
a chaotic behavior generated by {\it time-delay} differential equations \cite{fa}. 
A classic example of time-delay differential equation with chaotic solutions is the 
Mackey-Glass equation:
$$
\frac{du(t)}{dt} = \frac{0.2 \cdot u(t-\tau)}{(1+u(t-\tau)^{10})} - 0.1 \cdot u(t)  \eqno{(13)} 
$$
Figure 21 shows spectrum of a solution of this equation for the time-delay $\tau=30$. 
The dashed straight line indicates an exponential decay (cf. Fig. 20). 

In the Parker dynamo, which was generalized in the Refs. \cite{kle},\cite{zha},\cite{sokol}, 
a time-delay in the back influence of the magnetic field on the $\alpha$-effect \cite{bass},\cite{resh}
$$
\alpha (\vartheta,t,\tau)=\frac{\alpha_0(\vartheta)}{1+B^2(\vartheta,t-\tau)} \eqno{(14)}
$$
(where $B$ is the azimuthal component of the magnetic field) can significantly
change the evolution of the magnetic field even for a small time delay. In particularly, this 
non-linear delay can result in appearance of processes with periods much longer than 
the fundamental period through a parametric resonance \cite{resh}.

In the dynamo models that have physically distinct source layers the 
finite time is required in order to transport magnetic flux from one layer 
to another (a time-delay involved in the $\alpha$-quenching mechanism due to the
Lorentz feedback). Even the Duffing equations for $B$ can be obtained for the dynamo 
models using the delay idea in this case (cf. Eq. (14) in the Ref. \cite{ws}):
$$
\ddot{B} + \omega_0^2 B +\gamma \dot{B} + \beta B(t-\tau)f(B(t-\tau))= 0 \eqno{(15)}
$$
where $f(B(t-\tau))$ is a quenching factor, which can be approximated by a nonlinear
function. In particular, in the Ref. \cite{ws} (see also Ref. \cite{chau}) this function has been approximated as:
$$
f(B)=\frac{1}{4} (1+ \texttt{erf}~(B^2-B_{min}^2)(1-\texttt{erf}~(B^2-B_{min}^2)  \eqno{(16)}
$$
and for small $(B^2-B_{min}^2)$ we obtain a nonlinear delayed Duffing equation
$$
\ddot{B} + \Omega_0^2 B +\gamma \dot{B} + \beta' B^3(t-\tau)= 0 \eqno{(17)}
$$
where $\Omega_0^2$,$\gamma$ and $\beta'$ are certain constants. The subharmonic resonances and chaotic regimes 
are also known for the delay Duffing equations with a periodic forcing.
\begin{figure} \vspace{-3cm}\centering
\epsfig{width=.7\textwidth,file=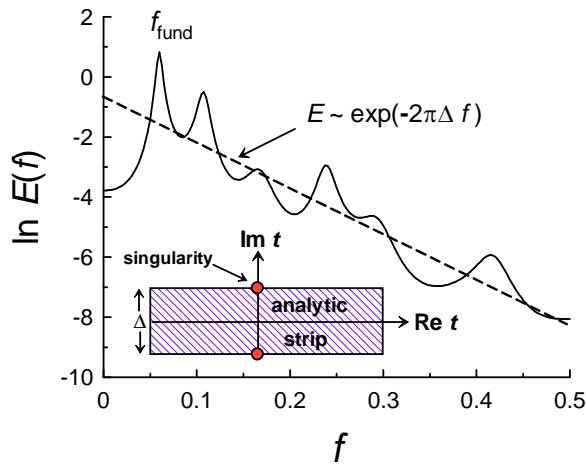} \vspace{-5cm}
\caption{Spectrum of the Mackey-Glass chaotic time-series. The dashed straight line 
indicates an exponential decay. The insert shows 
a sketch of corresponding complex-time plane.}
\end{figure}

It is also significant for those dynamo models that have 
spatially segregated source regions for the poloidal and toroidal magnetic field components 
(such as, for instance, the Babcock- Leighton dynamo mechanism \cite{chau}). In the global 
dynamo models that include meridional circulation the time delay related to the 
circulation should be comparable to global rotation period (see below).  

Nature of the exponential decay of the power spectra
of the chaotic systems (Figs. 20 and 21) is still an unsolved mathematical
problem. A progress in solution of this problem
has been achieved by the use of the analytical continuation
of the equations in the complex domain (see, for 
instance, \cite{fm}). In this approach the exponential decay
of chaotic spectrum is related to a singularity in the
plane of complex time, which lies nearest to the real axis (see 
the insert in Fig. 21).
Distance between this singularity and the real axis determines
the rate of the exponential decay. For many interesting cases 
chaotic solutions are analytic in a finite strip around the real time axis. 
This takes place, for instance for attractors bounded in the real 
domain (the Lorentz attractor, for instance). 
In this case the radius of convergence of the Taylor series 
is also bounded (uniformly) at any real time. 

Let us consider, for 
simplicity, solution $u(t)$ with simple poles only, and to define the Fourier 
transform as follows
$$
\tilde{u}(f) =(2\pi)^{-1/2} \int_{-T_e/2}^{T_e/2} dt~e^{-i 2\pi f t} u(t)  \eqno{(18)}
$$  
Then using the theorem of residues
$$
\tilde{u}(f) =i (2\pi)^{1/2} \sum_j R_j \exp (i 2\pi f x_j -|2\pi f y_j|)  \eqno{(19)}
$$
where $R_j$ are the poles residue and $x_j + iy_j$ are their location in the relevant half
plane, one obtains asymptotic behavior of the spectrum $E(f)= |\tilde{u}(f)|^2$ at large $f$
$$
E(f) \sim \exp (-4\pi |y_{min}|~ f)  \eqno{(20)}
$$
where $y_{min}$ is the imaginary part of the location of
the pole which lies nearest to the real axis. In the case of symmetric analytic strip with a width 
$\Delta= 2 |y_{min}|$:
$$
E(f) \sim \exp (-2\pi \Delta~f )  \eqno{(21)}
$$
(cf. the insert in Fig. 21). 

The chaotic spectrum provides two different characteristic
time-scales for the chaotic system: a period corresponding to
fundamental frequency of the system, $f_{fund}$, and a period
corresponding to the exponential decay rate, $2\pi \Delta$ 
(cf. Eq. (21)). The fundamental period can be estimated
using position of the low-frequency peak (cf. Figs. 20 and 21), while the
exponential decay rate period $2\pi \Delta$ can be estimated
using the slope of the straight line of the broad-band part
of the spectrum in the semi-logarithmic representation. In the case 
of the global solar dynamo the width of the analytic strip $\Delta$ 
can be theoretically estimated using the Carrington solar rotation period: 
$\Delta \simeq T_c \simeq 25.38$ days. 
This period roughly corresponds to the solar rotation at a latitude of 26 deg, 
which is consistent with the typical latitude of sunspots (cf. Fig. 20).  \\

Additionally to the exponential spectrum (Fig. 20), let us check the chaotic 
character of the wavelet regression detrended 
fluctuations calculating the largest Lyapunov exponent: $\lambda_{max}$. A strong indicator for the presence 
of chaos in the examined time series is condition $\lambda_{max} >0$. If this is the case, then 
we have so-called exponential instability. Namely, 
two arbitrary close trajectories of the system will diverge apart exponentially, that is 
the hallmark of chaos. To calculate $\lambda_{max}$ we used a direct algorithm developed by 
Wolf et al. \cite{w}. Figure 22 shows 
the pertaining average maximal Lyapunov exponent at the pertaining time, calculated for the data set shown in 
Fig. 18. The largest Lyapunov exponent converges very 
well to a positive value $\lambda_{max} \simeq 0.286~ mon^{-1} > 0$.\\
\begin{figure} \vspace{-2cm}\centering
\epsfig{width=.7\textwidth,file=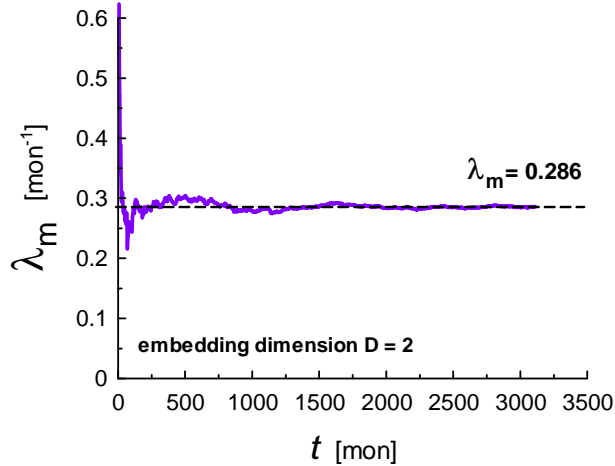} \vspace{-5.5cm}
\caption{The pertaining average maximal Lyapunov exponent 
at the pertaining time, calculated for the same data 
as those used for calculation of the spectrum (Figs. 19 and 20). 
The dashed straight line indicates convergence to a positive value.}
\end{figure}
\section{Flip-flop phenomenon}

It should be noted that the same period $\sim 3.7$ years was recently found for the so-called 
flip-flop phenomenon of the active longitudes in solar activity \cite{bu},\cite{bu2}. Sunspots 
are tend to pop up preferably in certain latitudinal domains and move toward the equator due to 
the 11-year cycle. Recently, strong indications of non-uniform {\it longitudinal} distribution of 
sunspots (active longitudes) was reported and analyzed in a dynamic frame related
to the mean latitude of sunspot formation, in which the active 
longitudes persist for the last eleven solar 11-years cycles 
(see Refs. \cite{bu},\cite{bu2} and references therein). At any given time, one of the two active 
longitudes (approximately $180^0$ apart) exhibits a stronger activity - dominance. Observed alternation 
of the active longitudes dominance in 3.7 years on average was called as flip-flop phenomenon 
\cite{bu}. It seems rather plausible that the observed flip-flop period and the fundamental period of the 
wavelet regression detrended fluctuations of solar activity (Fig. 20) have the same origin. In this vein, 
the observation \cite{bu},\cite{mh} that the period of the flip-flop phenomenon follows to 
variations of the real length of the sunspot cycle (which has the 11-years period on average only) 
supports the idea of the one-third subharmonic resonance as a background of the 11-years cycle of solar 
activity.

Another relevant example of the 3.7 years period appearance is the interplanetary
magnetic field polarity variations \cite{gg}. Close value ($\sim 3.5$ years) of a period
of the geomagnetic $aa$ index was reported in Ref. \cite{k}. \\

\section{Multidecadal coherence}

The question: Whether there is a coherence between solar activity and global temperature dynamics, 
was widely discussed in relation to the global warming. Naturally, in relation to the global warming 
just trends (solid curves in the Figs. 1 and 17) of the corresponding dynamical processes were studied. 
Now we can compare the wavelet regression detrended components of these processes (Figs. 2 and 18). 
Since the dominating periods of these detrended dynamical processes are different (3 and 3.7 years) 
we will look at a low-frequency domain (with frequency $f < 0.3y^{-1}$). The cross spectrum 
$E_{1,2}(f)$ of two processes $x_1(t)$ and $x_2(t)$ is defined 
by the Fourier transformation of the cross-correlation function normalized by the product of square root of the univariate power spectra $E_1(f)$ and $E_2(f)$:
$$
E_{1,2}(f)= \frac{\sum_{\tau} \langle x_1(t)x_2(t-\tau) \rangle 
\exp (-i2\pi f \tau)}{2\pi \sqrt{E_1(f)E_2(f)}} \eqno{(22)}
$$
the bracket $\langle... \rangle$ denotes the expectation value. The cross spectrum can be decomposed
into the phase spectrum $\phi_{1,2} (f)$ and the coherency $C_{1,2}(f)$:
$$
E_{1,2}(f)= C_{1,2}(f) e^{-i \phi_{1,2} (f)}  \eqno{(23)}
$$
Because of the normalization of the cross spectrum the coherency
is ranging from $C_{1,2}(f)=0$, i.e. no linear relationship between
$x_1(t)$ and $x_2(t)$ at $f$, to $C_{1,2}(f)=1$, i.e. perfect linear relationship. 
\begin{figure} \vspace{-1cm}\centering
\epsfig{width=.7\textwidth,file=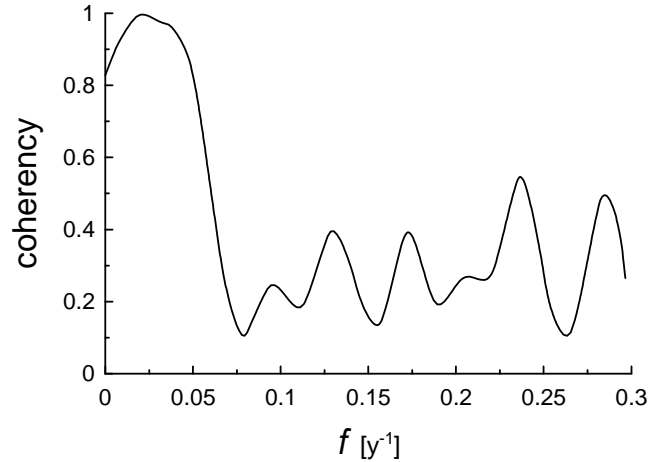} \vspace{-6cm}
\caption{Coherency between the wavelet regression detrended solar activity and global temperature 
in a low-frequency domain.}
\end{figure}
 
Figure 23 shows the coherency between the wavelet detrended solar activity and global temperature 
(for the period 1880-2009yy) in the low-frequency domain, computed using the fast Fourier transformation. 
One can see a very high coherency for the time periods larger than 20 years. Let us recall that 
the 22y period corresponds to the Sun's magnetic poles polarity switching and can be a base for a multidecadal 
chaotic coherence between solar activity and global climate (cf. Refs. \cite{b}). It is knowm that there is no clear correlation between the data dominated by the trends (shown in the Figs. 1 and 17). Therefore, one can conclude 
that just the trends are presumably not correlated in this case, whereas the large-scale fluctuations (wavelet regression detrended: Figs. 2 and 18) exhibit a very high coherency for the time periods larger than 20 years. Thus, just the nonlinear nature of the Sun-climate interaction is the main cause of the well known difficulties 
in estalishing and investigation of this interaction.

\section{Conclusion}

In the considered complex natural systems the fluctuations are often of the same order as the trend itself. 
This phenomenon is related to a nonlinear character of the fluctuations. Has this nonlinear phenomenon 
an universal nature? The main three points of the above consideration are:

a) the known (crude) models can be reduced to the Duffing oscillator, 

b) the one-third subharmonic resonance is a generic property of this oscillator, 

c) the data indicates crucial role of this resonance in generation of the large-scale fluctuation. 

The next generation of the models of global climate should take into account the highly nonlinear nature of the Sun-climate interaction, which is not realized directly through the trends but through the large-scale fluctations 
(both for monthly and for multidecadal time scales).

At this stage of the data accumulation and analysis limitations of this conclusions are unknown. 
However, we believe that the above considered results could attract attention of the nonlinear scientists to the natural systems. Because, despite of their overwhelming complexity, these systems are governed by the simple physical laws.   \\

I thank to S.I. Abarzhi and to K.R. Sreenivasan for suggestions, criticism and encouragement.
The data were provided by National Climatic Data Center at NOAA, by Joint Institute for the Study
of the Atmosphere and Ocean, and by SIDC-team, World Data Center
for the Sunspot Index, Royal Observatory of Belgium. I also acknowledge that a software provided 
by K. Yoshioka was used at the computations.


\newpage

\begin{thebibliography}{99}
\bibitem{sal2} B. Saltzman, A. Sutera, and A. Hansen, A possible marine mechanism for internally 
generated long-period climate cycles, J. Atmos. Sci., {\bf 39}, 2634-2637 (1982).
\bibitem{nic} C. Nicolis, Long-term climatic transitions and stochastic resonance, J. Stat. Phys., {\bf 70}, 
3-14 (1993). 
\bibitem{jin} F.-F. Jin, F.-F. (1997), An equatorial ocean recharge paradigm for ENSO. Part I:
Conceptual model, J. Atmos. Sci., {\bf 54}, 811-829 (1997).
\bibitem{wl} W. B. White and Z. Liu, Non-linear alignment of El Nino to the 11-yr solar cycle, Geophys. Res. Lett. 
{\bf 35}, L19607, (2008). 
\bibitem{ot} E. Ott, Chaos in Dynamical Systems (Cambridge University Press, 2002).
\bibitem{ph} D. Permann and I. Hamilton, Wavelet analysis of time series for the Duffing oscillator: 
The detection of order within chaos, Phys. Rev. Lett., {\bf 69}, 2607 (1992).
\bibitem{b} A. Bershadskii, Chaotic climate response to long-term solar forcing variability, EPL
(Europhys. Lett.), {\bf 88}, 60004 (2009).
\bibitem{nm} A.H. Nayfeh and D.T. Mook, "Nonlinear Oscillations" (John Wiley \& Sons, 
a Wiley-Interscience Publication, 1979).
\bibitem{tw} S.M. Tobias, and N.O. Weiss, Resonant Interactions between Solar Activity and Climate, 
J. Climate, {\bf 13}, 3745 (2000). 
\bibitem{bkb} J. Brindley, T. Kapitaniak, and A. Barcilon, 
Chaos and noisy periodicity in forced ocean-atmosphere models, Phys. Lett. A, {\bf 167}, 179-184 (1992). 
\bibitem{nl} Yu.I. Neimark and P.S. Landa, Stochastic and Chaotic Oscillations,
(Dordrecht, Kluwer, 1992).
\bibitem{mj} S. Minobe, and F-f Jin., Generation of interannual and interdecadal climate oscillations
through nonlinear subharmonic resonance in delayed oscillators, Geophys. Res. Lett., {\bf 31}, 
L16206, (2004).
\bibitem{s} T.M. Smith, et al., Improvements to NOAA's Historical Merged Land-Ocean Surface Temperature Analysis (1880-2006), J. Climate, {\bf 21}, 2283 (2008).
\bibitem{ka} K. Kawamura et al., Northern Hemisphere forcing of climatic cycles in Antarctica over
the past 360,000 years,  Nature, {\bf 448}, 912-916 (2007).
\bibitem{sw} N. Scafetta, and B. J. West, Phenomenological solar contribution to
the 1900-2000 global surface warming, Geophys. Res. Lett., {\bf 33}, L05708 (2006).
\bibitem{o} T. Ogden, Essential Wavelets for Statistical Applications and Data Analysis 
(Birkhauser, Basel, 1997).
\bibitem{pl} P.S. Polito, and W.T. Liu, Global characterization of Rossby waves at several 
spectral bands, J. Geophys. Res. - Oceans, {\bf 108}, 3018 (2003).
\bibitem{zw} X. Zang and C. Wunsch, Spectral description
of low frequency oceanic variability, J. Phys. Oceanogr., {\bf 31} 3073 (2001).
\bibitem{s2} T.M. Smith, et al., Reconstruction of historical sea surface temperatures using 
empirical orthogonal functions, J. Climate, {\bf 9}, 1403 (1996).
\bibitem{zfw} X. Zang, L.L. Fu, and C. Wunsch, Observed reflectivity of
the western boundary of the equatorial Pacific Ocean. J. Geophys.
Res., {\bf 107}, 3150 (2002).
\bibitem{tzi} E. Tziperman, H. Scher, S.E. Zebiak and M. A. Cane, Controlling Spatiotemporal 
Chaos in a Realistic El Nino Prediction Model, Phys. Rev. Lett., {\bf 79}, 1034-37 (1997).
\bibitem{rh} M. Raymo and P. Huybers, Unlocking the mysteries of the ice ages, 
Nature {\bf 451}, 284-285 (2008)
\bibitem{sal} B. Saltzman, Dynamical paleoclimatology : generalized theory of global
climate change. (Academic Press, San Diego, 2001).
\bibitem{berg1} A. Berger, X. Li, and M.F. Loutre, Modelling northern hemisphere ice volume over
the last 3 Ma. Quaternary Science Reviews, {\bf 18}, 1-11 (1999).
\bibitem{ru} W.F. Ruddiman, Orbital insolation, ice volume, and greenhouse gases.
Quaternary Science Reviews, {\bf 22}, 1597-1629 (2003).
\bibitem{clar} P. Clark, D. Archer, D. Pollard, J. Blum, J., et al., 
The middle Pleistocene transition: characteristics, mechanisms, and implications for longterm
changes in atmospheric $CO_2$, Quaternary Sci. Rev., {\bf 25}, 3150-3184 (2006).
\bibitem{b3} A. Bershadskii, Journal of Cosmology, Nonlinear and Chaotic Ice Ages: Data vs. Speculations, 
{\bf 8}, 1893-1905 (2010).
\bibitem{vostok1} J.R. Petit, J.R., et al., Vostok Ice Core Data for 420,000 Years, IGBP PAGES/World Data Center 
for Paleoclimatology Data Contribution Series \#2001-076.  NOAA/NGDC Paleoclimatology Program, Boulder CO, USA (2001). 
\bibitem{vostok2} J. Jouzel J., et al., Climate and Atmospheric History of the Past 420,000 years from the 
Vostok Ice Core, Antarctica, Nature, {\bf 399}, 429-436 (1999).
\bibitem{blm} A. Berger, M.F. Loutre, and J. L. Melice, Equatorial insolation: from precession
harmonics to eccentricity frequencies, Clim. Past Discuss., {\bf 2}, 519-533 (2006).
\bibitem{hag} T.K. Hagelberg, G. Bond, and P. de Menocal, Milankovitch band forcing of sub-Milankovitch
climate variability during the Pleistocene, Paleoceanography, {\bf 9}, 545-558 (1994).
\bibitem{my} A. S. Monin and A. M. Yaglom, Statistical Fluid Mechanics, Vol. II (MIT
Press, Cambridge, 1975).
\bibitem{b1} A. Bershadskii, Multiscaling of Galactic Cosmic Ray Flux, 
Phys. Rev. Lett., {\bf 90}, 041101 (2003).
\bibitem{gibson} C.H Gibson, Kolmogorov Similarity Hypotheses for Scalar Fields:
Sampling Intermittent Turbulent Mixing in the Ocean and Galaxy, Proc. Roy. Soc. Lond. {\bf 434}, 149 (1991).
\bibitem{pet} A. Petrosyan, A. Balogh, M. L. Goldstein, et al., Turbulence in the Solar Atmosphere and Solar Wind, 
Space Science Reviews, {\bf 156}, 135-238 (2010).
\bibitem{l} A. Lazarian, Obtaining Spectra of Turbulent Velocity from Observations, Space Science Reviews,
{\bf 143}, 357-385 (2009).
\bibitem{lp} Lazarian, A. and D. Pogosyan: Velocity modification of HI power spectrum,
ApJ, {\bf 537}, 720L (2000).
\bibitem{clv} J. Cho, A. Lazarian, and E.T. Vishniac, Simulations of MHD Turbulence in a Strongly Magnetized 
Medium, Astrophys. J. {\bf 564}, 291-301  (2002) (see also arXiv:astro-ph/0205286).
\bibitem{sd} K.R. Sreenivasan and B. Dhruva, Is there scaling in high-Reynolds-number
turbulence? Prog. Theor. Phys. Suppl., {\bf 130}, 103-120 (1998).
\bibitem{g2} C.H. Gibson, R.N. Keeler, V.G. Bondur, et al., Submerged turbulence detection with optical 
satellites, Proc. of SPIE, {\bf 6680}-33 (2007).
\bibitem{sf} H. Svensmark and E. Friis-Christensen, Variation of cosmic ray
flux and global cloud coverage – A missing link in solar-climate
relationships, J. Atm. Sol. Terr. Phys., {\bf 59}, 1225-1232 (1997).
\bibitem{sh}  H. Svensmark, Influence of Cosmic Rays on Earth's Climate, Phys. Rev. Lett., {\bf 81}, 5027 (1998).
\bibitem{zf} G.P. Zank, and P.C. Frisch, Consequences of a Change in the Galactic Environment of the Sun,
The Astrophysical Journal, {\bf 518}, 965-973 (1999).
\bibitem{ms} N. Marsh and H. Svensmark, Low Cloud Properties Influenced by Cosmic Rays, Phys. Rev. Lett.,
{\bf 85}, 5004 (2000).
\bibitem{shav1} N. Shaviv, Cosmic Ray Diffusion from the Galactic Spiral Arms, Iron Meteorites, and a Possible Climatic Connection, Phys. Rev. Lett. {\bf 89}, 051102 (2002).
\bibitem{shav2} N. Shaviv, The spiral structure of the Milky Way, cosmic rays, and ice age epochs on Earth, New Astronomy, {\bf 8,} 39 (2003).
\bibitem{ch} M. Christl et al., Evidence for a link between the flux of galactic cosmic rays and Earth's climate during the past 200,000 years,   Journal of Atmospheric and Solar-Terrestrial Physics, {\bf 66}, 313-322 (2004).
\bibitem{kaz} J. Kazil, R.G. Harrison, and E.R. Lovejoy,
Aerosol nucleation over oceans and the role of galactic cosmic rays, Atmos.
Chem. Phys., {\bf 6}, 4905-4924 (2006).
\bibitem{h} M. Haverkorn et al., The outer scale of turbulence in the magnetoionized galactic interstellar medium, 
ApJ, {\bf 680} 362-370 (2008).
\bibitem{du} J. Duplissy et al. Results from the CERN pilot CLOUD experiment, Atmos. Chem. Phys., {\bf 10},
1635-1647 (2010).
\bibitem{kle} N. Kleeorin, K. Kuzanyan D.Moss, et al., Magnetic helicity evolution during the solar activity cycle:
observations and dynamo theory, A\&A, {\bf 409}, 1097-1107 (2003).
\bibitem{zha} H. Zhang, D. Sokoloff, I. Rogachevskii, et al., The radial distribution of magnetic helicity in the solar
convective zone: observations and dynamo theory, Mon. Not. Roy. Astron. Soc., {\bf 365} 276-288 (2006).
\bibitem{sokol} D. Sokoloff, S.D. Bao, N. Kleeorin, et al., The distribution of current helicity at the solar surface at the beginning of the solar cycle, Astron. Nach., {\bf 327}, 876-884 (2006).
\bibitem{b4} A. Bershadskii, EPL (Europhys. Lett.), Transitional dynamics of the solar convection zone, 
{\bf 85} 49002 (2009).
\bibitem{pn} M. Palus, and D. Novotna, Sunspot Cycle: 
Phys. Rev. Lett., {\bf 83}, 3406 (1999).
\bibitem{mini} P.O. Mininni, D.O. Gomez, and G.B. Mindlin, Simple Model of a Stochastically Excited Solar Dynamo, 
Solar Phys., {\bf 201}, 203-223 (2001).
\bibitem{frick} P. Frick, D. Galyagin, D. Hoyt et al., Wavelet analysis of solar activity 
recorded by susspots grpups, A\&A, {\bf 328}, 670-681 (1997).
\bibitem{sokol2} D. Sokoloff, The Maunder Minimum and the Solar Dynamo, Solar Physics, {\bf 224}, 145-152 (2004).
\bibitem{nocera}  L. Nocera, Subharmonic oscillations of a forced hydromagnetic cavity, 
Geoph. \& Astroph. Fluid Dynamics, {\bf 76}, (1994) 239.  
\bibitem{plo} D. Passos and I. Lopes, A Low-Order Solar Dynamo Model: Inferred Meridional Circulation Variations Since 1750, ApJ, {\bf 686}, 1420 (2008). 
\bibitem{oht} N. Ohtomo,  K. Tokiwano, Y. Tanakaet. et al., Exponential Characteristics of 
Power Spectral Densities Caused by Chaotic Phenomena, J. Phys. Soc. Jpn., {\bf 64}, 1104-1113 (1995). 
\bibitem{fa} J. D. Farmer, Chaotic attractors of an infinite-dimensional dynamical system, 
Physica D, {\bf 4}, 366-393 (1982).
\bibitem{sig} D.E. Sigeti, Survival of deterministic dynamics in the presence of 
noise and the exponential decay of power spectra at high frequency, Phys. Rev. E, {\bf 52}, 2443-2457 (1995).
\bibitem{fm} U. Frisch and R. Morf, Intermittency in nonlinear dynamics and singularities at complex times
Phys. Rev., {\bf 23}, 2673-2705 (1981).
\bibitem{bass} A.P. Bassom, K.M. Kuzanyan, and A. M. Soward, A nonlinear dynamo wave riding 
on a spatially varying background, Proc. R. Soc. London A, {\bf 455}, 1443-1481 (1999).
\bibitem{resh} M.Y. Reshetnyak, Astronomy Reports, Time-Lag Effects in Parker's Dynamo, {\bf 54}, 1047-1052 (2010).
\bibitem{ws} A.L. Wilmot-Smith, D. Nandyet, G. Hornig and P.C.H. Martens, A time delay model for 
solar and stellar dynamos, ApJ, {\bf 652}, 696-708 (2006).
\bibitem{chau} P. Charbonneau, C. St-Jean, and P. Zacharias, Fluctuations in Babcock-Leighton Dynamos. I. Period Doubling and Transition to Chaos, ApJ, {\bf 619} 613-622 (2005).
\bibitem{w} A. Wolf, J.B. Swift, H.L. Swinney and J.A. Vastano, Determining Lyapunov exponents from a time series, Physica D, {\bf 16}, 285-317 (1985).
\bibitem{bu} S.V. Berdyugina and I.G. Usoskin, Active longitudes in sunspot activity: Century scale persistence, 
A\&A {\bf 405}, 1121-1128 (2003).
\bibitem{bu2}I.G. Usoskin, S.V. Berdyugina, D. Moss, and D.D. Sokoloff, Long-term persistence of solar active longitudes and its implications for the solar dynamo theory, Advances in Space Res., {\bf 40}, 951-958 (2007).
\bibitem{mh} K. Mursula, and T. Hiltula, Systematically Asymmetric Heliospheric Magnetic Field: Evidence for a Quadrupole Mode and Non-Axisymmetry with Polarity Flip-Flops, 
Solar Phys. {\bf 224}, 133-143 (2004).
\bibitem{gg} A.L.C. Gonzalez and W.D. Gonzalez, Periodicities in the Interplanetary Magnetic 
Field Polarity, J. Geophys. Res., {\bf 92}, 4357-4375 (1987).
\bibitem{k} R.P. Kane, Quasi-biennial and quasi-triennial oscillations in geomagnetic activity indices, Ann. Geophys., 
{\bf 15}, 1581-1594 (1997).
\end{thebibliography}
\end{document}